\documentclass[10pt,twocolumn,aps,reprint,prl]{revtex4-1}
\usepackage{multirow}
\usepackage{graphicx}

\begin{document}

\title{Erosion dynamics of a wet granular medium}

\author{Gautier Lefebvre}

\email{gautier.lefebvre@saint-gobain.com}

\affiliation{Surface du Verre et Interfaces, UMR 125 CNRS/Saint-Gobain, 39, quai
Lucien Lefranc, F-93303 Aubervilliers, Cedex, France.}

\author{Pierre Jop}

\email{pierre.jop@saint-gobain.com}

\homepage{www.svi.cnrs-bellevue.fr}

\affiliation{Surface du Verre et Interfaces, UMR 125 CNRS/Saint-Gobain, 39, quai
Lucien Lefranc, F-93303 Aubervilliers, Cedex, France.}
\begin{abstract}
Liquid may give strong cohesion properties to a granular medium, and
confer a solid-like behavior. We study the erosion of a fixed circular
aggregate of wet granular matter subjected to a flow of dry grains
inside a half-filled rotating drum. During the rotation, the dry grains
flow around the fixed obstacle. We show that its diameter decreases
linearly with time for low liquid content, as wet grains are pulled-out
of the aggregate. This erosion phenomenon is governed by the properties
of the liquids. The erosion rate decreases exponentially with the
surface tension while it depends on the viscosity to the power -1.
We propose a model based on the force fluctuations arising inside
the flow, explaining both dependencies: the capillary force acts as
a threshold and the viscosity controls the erosion time scale. We
also provide experiments using different flowing grains confirming
our model. 
\end{abstract}
\maketitle

\section{Introduction}

It is commonly known that the addition of a small amount of liquid
in a granular medium brings cohesion properties due to the surface
tension of the liquid. Such a mixture may have a strong solid-like
behavior \cite{Herminghaus,Mitarai}, and for instance enables to
build sand castles. Properties and rheology of homogeneous wet granular
materials have received a lot attention from experimental \cite{Iveson,Moller}
and numerical point of view \cite{Rognon,Richefeu}. However the situations
encountered in nature or industry often present heterogeneous systems,
where the liquid content is not homogeneously distributed over space.

This is the case for some landslides where the basal material is more
cohesive than the flowing one. Such a situation arises for example
because of humidity. To model their dynamics, the evolution of the
interface between the erodible ground and the flowing material is
still studied experimentally or numerically \cite{Mangeney,Iverson}
and the effect of the cohesion on erosion remains unknown.

In the industrial context, many processes blend powders and grains
with liquids. Understanding the mechanisms of the spreading of the
liquid is important to avoid lump formation when preparing dough in
food industry, but also in the granulation phenomenon to obtain pills
in pharmaceutical industry \cite{Ennis}, or during the production
of slurries for mortar or concrete in building materials \cite{Cazacliu,Collet}.
During the first stages of the blending, wet areas are in contact
with dry flowing grains.

We can expect morphological evolution of the cohesive medium through
exchanges between the two areas. More precisely, in the case of low
water content, one may expect erosion of the cohesive phase to occur
by extraction of grains from the cohesive medium. Despite the large
interest of industry in these processes, the precise mechanisms of
these initial steps are not known. What are the exchange rates between
these phases?

\begin{figure}
\includegraphics[scale=0.23]{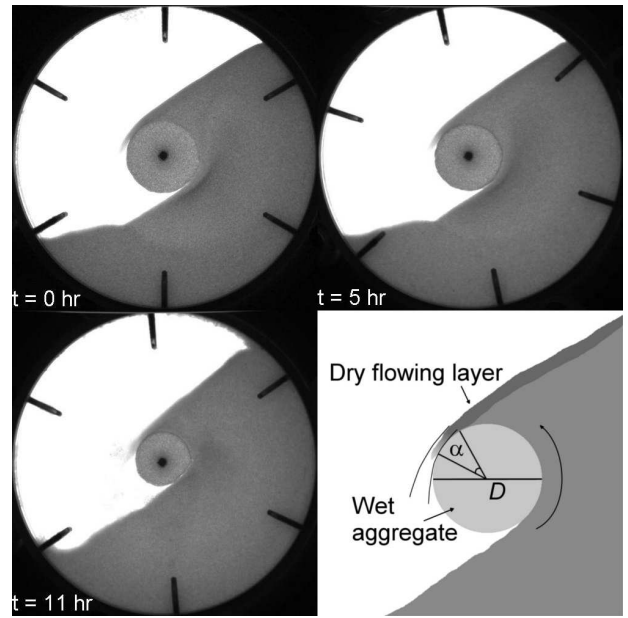}

\caption{\label{fig:drum view}Images of the drum during the experiment, with
glass beads and silicone oil V100. We can see the evolution of the
aggregate diameter. The dark spot in the middle of the cylinder is
a glued tiny pillar helping to keep the aggregate at the center. On
the bottom right is a sketch of the region of interest. $D$ is the
diameter of the aggregate. $\alpha$ is the portion submitted to erosion. }
\end{figure}

Although erosion of granular bed by a liquid flow is well described
by transport models \cite{Charru,Lajeunesse}, erosion by a granular
flow is not as well understood yet. Nonetheless, attempts to model
the effect of the flowing grains on erosion are found in different
fields in literature. Erosion of a substrate by a granular phase is
of interest for geomorphology, where empirical laws are derived from
field observation and from model experiments \cite{Hsu,Sklar04}.
The erosion rate is usually related to the kinetic energy of impacting
grains \cite{Sklar04} like in the seminal models of wear production
by sand blasting \cite{Bitter,Finnie}, where the erosion rate scales
with the velocity of impacting grains to a power between 2 and 5 \cite{Meng}.
In the case of enduring contact, fretting wear has been shown to be
proportional to the normal load \cite{Archard}. However two main
points are questionable: first, the stress and flowing conditions
at the granular interface are still matter of debate, specially for
dense flows \cite{Artoni09} and are hardly linked to the erosion
processes \cite{Fillot,Hsu}. What is the driving mechanism for granular
erosion? Second, the cohesive media may not be considered as a continuous
material since its internal length scale is of the same order of size
as the flowing grain. Thus the previous laws derived for brittle or
plastic materials showing that the erosion rate decreases with the
square of the tensile strength may not apply \cite{Sklar01}. We expect
on the opposite that forces developed by a stretched capillary bridge
\cite{Herminghaus} will govern the erosion process.

In this paper, we study this erosion phenomenon experimentally with
a model system. Here, cohesion is brought by capillary bridges only.
The eroding flow is constituted of dry-dense granular matter. We explore
the dynamics of erosion in regard of the properties of the liquids
in the wet granular medium, and of the properties of the flowing grains.
We present a model of erosion that reveals unexpected dependencies
and that may be used to better understand the mixing issues mentioned
above.

\section{Experimental set-up}

\subsection{Materials and device}

We used a thin Plexiglas cylinder with an inner diameter of 14.2 cm
and a depth of 0.5 cm as rotating drum (Fig. \ref{fig:drum view}).
These dimensions correspond to an aspect ratio around 28, so the drum
can be considered as 2-dimensional. We introduced wet material, of
controlled liquid content, to form a circular aggregate of diameter
$D\simeq3$ cm at the center of the drum. To do so, we first mix an
amount of grains and liquid, with the help of a spatula, until we
obtain a homogeneous mixture. The cohesion forces that ties the aggregate
will also make it stick to the vertical drum walls, and remain at
the center. Then, we filled half the place left with dry grains before
closing the drum, and putting it into rotation. Different types of
material have been employed, and are summarized in table \ref{tab:Granular-materials-employed}.
Phonolite has an important roughness, and thus is close to real materials.
We also used several kinds of beads ranging from 0.2 to 1.3 mm in
diameter. We will use different notations for the flowing-grain diameter
$d$ and for the radius of wet grains $r$ inside the aggregate. Different
liquids were used, water, glycerol, ethylene glycol and silicone oils,
in order to vary the surface tension $\gamma$, from 20 to 70 mN/m,
and viscosity $\eta$, from 1 to $10^{4}$ mPa.s.

For the range of measured surface tension $\gamma$, the typical granular
Bond number is always high, with $Bo_{g}=2\pi\gamma r/mg>50$, where
$r$ is the wet grain radius and $m$ the mass of the grain. That
is why cohesion from capillary bridges here can easily overcome the
gravity. The aggregate has then enough cohesion to sustain itself,
and not break under its own weight. The capillary length $l_{c}=\sqrt{\gamma/\rho_{liq}g}$
is around 2.6 mm for water, much larger than the typical size of capillary
bridges, so gravity will not deform them. Finally, drainage is limited
by viscous effects, as the liquid should travel through thin films
of liquids, whose thickness is the roughness of the bead surface,
$\delta$. Considering the hydrostatic pressure on the size of the
aggregate, we obtain a drainage time on a distance $r$ of $t_{drain}=\eta D/\delta^{2}\rho_{liq}g\simeq10^{4}$
s for $\eta=$10 mPa.s. This means that gravity driven drainage effects
are prevented, as the rotation period is much smaller than this time.

\begin{table}
\begin{tabular}{|c|c|c|}
\hline 
Granular material  & Density $\rho$ (g/cm$^{3}$) & Size $d$ (mm)\tabularnewline
\hline 
Phonolite grains  & 2.6  & 0.8-1\tabularnewline
\hline 
\multirow{4}{*}{Glass Beads (GB)} & \multirow{4}{*}{2.5 } & 0.2-0.4\tabularnewline
 &  & 0.5\tabularnewline
 &  & 0.8-1\tabularnewline
 &  & 1-1.3\tabularnewline
\hline 
Polystyrene Beads (PS) & 1  & 0.5\tabularnewline
\hline 
Zirconium Silicate Beads  & 3.8  & 0.5\tabularnewline
\hline 
Zirconium Oxyde Beads  & 5.5  & 0.5\tabularnewline
\hline 
Stainless Steel Beads  & 7.9  & 0.5\tabularnewline
\hline 
\end{tabular}

\caption{Granular materials employed in the experiments, and physical properties.
The flowing-grain diameter is denoted by $d$ and the radius of grains
inside the aggregate is denoted by $r$. \label{tab:Granular-materials-employed}}
\end{table}

\begin{table}
\begin{tabular}{|c|c|c|c|}
\hline 
Liquids & Viscosity  & Surface Tension & Contact Angle\tabularnewline
 & (mPa.s)  & (mN/m) & \tabularnewline
\hline 
Water  & 1.0  & 72 & 32\tabularnewline
\hline 
Silicone Oils  & \multirow{2}{*}{10-10000} & \multirow{2}{*}{21} & \multirow{2}{*}{<5}\tabularnewline
V10-V10000  &  &  & \tabularnewline
\hline 
Water-Glycerol  & 1.3-560  & 60-72 & 30-46\tabularnewline
\hline 
Water- & \multirow{2}{*}{10-17 } & \multirow{2}{*}{48-49} & \multirow{2}{*}{31}\tabularnewline
Ethylene Glycol &  &  & \tabularnewline
\hline 
\end{tabular}

\caption{Liquids employed in the experiments, and the measured physical properties.
\label{tab:Liquids-employed}}
\end{table}

\subsection{Measurements}

During the rotation, snapshots are regularly taken, and the relevant
information is retrieved by image analysis. The dry grains and the
flow can be identified from grayscale levels, and the shape of the
aggregate can be followed during the experiment. The lateral view
of the drum of Fig. \ref{fig:drum view} shows how the flow is modified
by the aggregate. Most of the grains flow above it, but a small part
of it can pass under. Meanwhile, moisture and temperature of the room
are recorded in the process. Temperature is needed for a precise determination
of the liquid properties, like viscosity. Humidity is a source of
liquid in the granular medium, as shown by Bocquet \textit{et al.}
in \cite{Bocquet}, which can influence the results. However, we checked
that the influence of this parameter is weak compared to our data
dispersion.

For experiments with glass beads, the progressive liquid spreading
due to erosion in the dry area can alter the flow properties. Cohesion
appears in the surrounding medium until it sticks to the edges of
the drum. The setting is then widely modified, so we use only the
first part of the experiment, when the spreading of the liquid is
still low enough.

\section{Experimental results}

\subsection{Multiple regimes}

\begin{figure}
\includegraphics[scale=0.5]{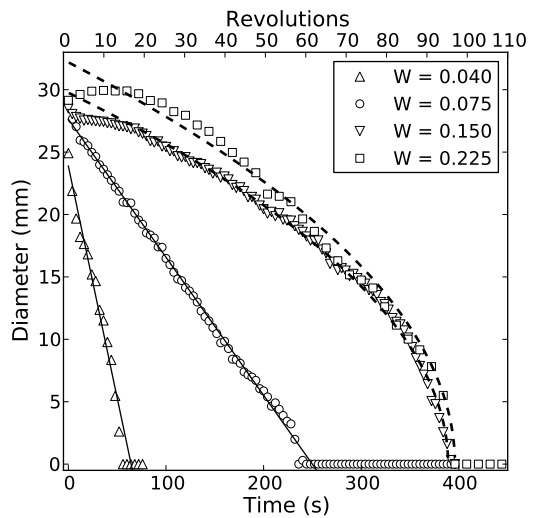}\caption{Time evolution of the aggregate's diameter, with phonolite grains,
for different water contents. The upper axis is the number of revolutions
of the drum. Plain lines are linear fits, dashed lines are square
root functions fits.\label{fig:erosion-phono}}
\end{figure}

Trials have first been led with phonolite and water, for different
liquid contents. In the following, we choose to use $W$, the ratio
of liquid volume on the total volume of aggregate. As there is no
compaction of the aggregates in our experiments, $W$ is not changed
without addition or withdrawal of liquid. We had to used quite high
levels of liquid contents due to the high roughness of phonolithe
grains, between 4 and 22.5\%. The rotation speed was maintained at
0.349 rad/s, corresponding to an 18 s period. Figure \ref{fig:erosion-phono}
shows the variation of the aggregate diameter, from its initial value
to zero, when it disappears in the flow. The lifetime of the aggregate
(a few minutes) increases with $W$. For low liquid contents, the
diameter decreases linearly with time. Such behavior can easily be
justified by simple assumptions. In this case, capillary bridges are
individual, so the behavior of interfacial grains does not depend
on the other grains farther below the surface. The erosion of the
grains is then a local mechanism. Moreover, the flow properties are
stationary, thus we assume that the mechanical action of the flow
is constant with time. Since the wet aggregate is assumed to be homogeneous,
the local erosion rate should also be constant with time. Finally,
the diminution of the aggregate area should scale with the portion
$\alpha D/2$ of perimeter undergoing the erosion (Fig \ref{fig:drum view}),
leading to $-\frac{d(\pi D^{2})}{dt}\propto\alpha D$. We assume that
the angle $\alpha$ depends on the geometry only, and remains unchanged.
The diameter decrease would then be linear: $\frac{dD}{dt}=cst$. 

For the higher liquid contents, we have a slower evolution, and even
a growth phase for $W$=22.5\%. This growth is probably enabled by
the rearrangement of the liquid distribution. If the liquid network
is sufficiently connected, suction may bring the liquid to the edge
of the aggregate and create new capillary bridges, with initially
dry grains, coming from the flow. We saw that gravity-driven drainage
was not permitted because of viscous effects. Moreover, when the liquid
content is high enough, the liquid is not distributed anymore into
single capillary bridges. The Laplace pressure may be weaker in water
pockets implying more than two grains, and on the opposite, it will
plays fully on the newly formed bridge. That is why liquid transport
may occur here, as it is driven by capillary effects, stronger than
gravity at the grain scale. The diameter seems to follow a square
root collapse with time (Fig. \ref{fig:erosion-phono}). As a simple
tentative explanation, we can consider that the flowing grains pump
the liquid toward the surface of the aggregate, thus the erosion will
be lower at the beginning while the degree of saturation in the core
decreases. Then, the erosion rate increases when the diameter decreases,
as the water content is lower inside. We reserve this issue for future
work.

The linear regime appears then as a simpler process, which does not
involve liquid migration inside the aggregate. In this regime, we
can measure an erosion rate $E$ from the slope of the lines, to quantify
the speed of the process. We define $E$ as a dimensionless parameter,
by rescaling the aggregate diameter $D$ by the beads diameter $d$,
and the time by the rotation period of the drum $T$: 
\begin{equation}
\frac{D-D_{0}}{d}=-E\frac{t}{T}.\label{eq: Definition Erosion rate}
\end{equation}
$E$ is then a positive number, counting the number of layers of grains
eroded from the surface of the aggregate, for each revolution of the
drum.

\subsection{Liquid properties}

\begin{figure}
\includegraphics[scale=0.5]{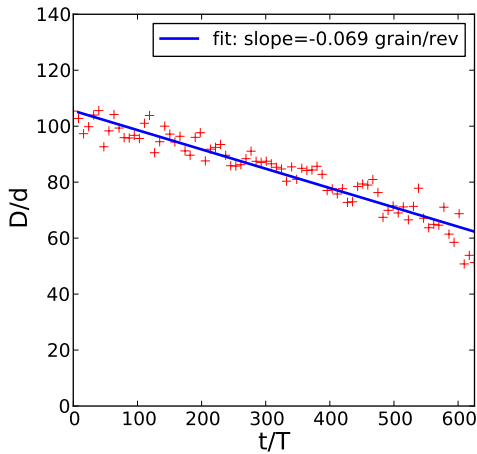}\caption{Linear decrease of the aggregate diameter with 200-400 $\mu$m
glass beads, for $W$=0.3\%. The liquid used is silicone oil V100.
\label{fig:Linear-decrease}}
\end{figure}

In the following of the paper, we use relatively low liquid contents
to remain in the linear regime of erosion, for which we can measure
an erosion rate. The typical value will be $W=0.3\%$, and otherwise
precised. Figure \ref{fig:Linear-decrease} shows a typical run of
the experiment. The diameter of the aggregate decreases linearly,
and the fit provides a value of erosion rate $E$ as defined in equation
(\ref{eq: Definition Erosion rate}).

\begin{figure}
\includegraphics[scale=0.5]{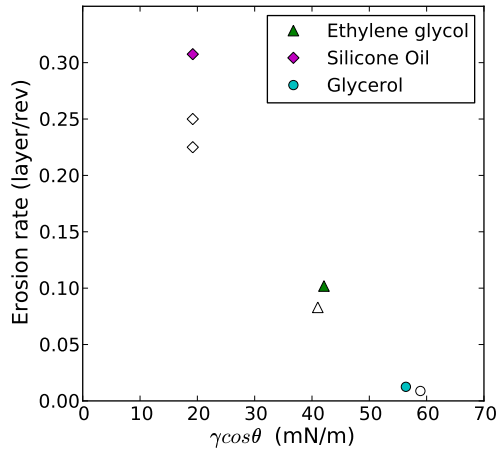}\caption{Erosion rate versus $\gamma\cos\theta$ for different liquids and
glass beads. For filled symbols viscosity is around 10 mPa.s, and
for open ones around 20 mPa.s\label{fig:gamma}}
\end{figure}

To stretch a capillary bridge linking two grains, one must overcome
two forces, one coming from the surface tension, and the other one
from the viscosity of the liquid. The forces developed by a capillary
bridge are approximated by \cite{Herminghaus}: 
\begin{equation}
F_{cap}=2\pi\gamma r\cos\theta,\hspace{1em}F_{visc}=\frac{3}{2}\pi r^{2}\eta\frac{1}{s}\frac{ds}{dt},\label{eq:cohesion forces}
\end{equation}
$\theta$ being the contact angle, and $s$ the separation distance
between the two beads. These are first order approximations regarding
$s$. Only the normal viscous dissipation is considered, and we assume
that most of the viscous dissipation arises from normal displacement.
We expect then both surface tension and viscosity to increase the
resistance to erosion, and thus to observe lower erosion rates when
they increase. In order to observe the effect, an other set of experiments
have been carried out with various liquids. We used the same protocol
as previously, with a 24 s period. The liquid content is fixed at
$W$=0.3\% or otherwise precised. Figure \ref{fig:gamma} shows the
evolution of the erosion rate with the surface tension. The different
liquids have similar viscosity, from 10 to 20 mPa.s. The erosion rate
strongly decreases even for a small range of surface tension.

\begin{figure}
\includegraphics[scale=0.5]{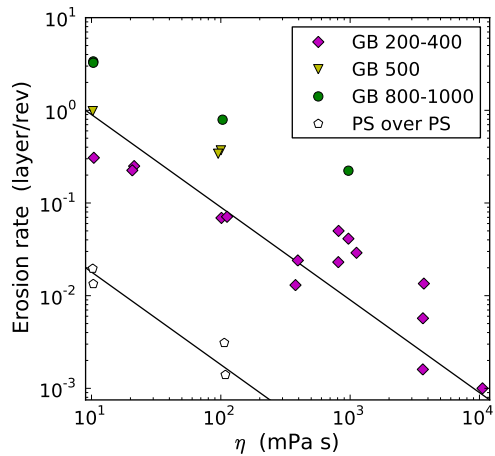}\caption{Erosion rate versus viscosity for different beads. Surface tension
is around 21 mN/m. In each case, we made the aggregate with the same
200-400 $\mu$m glass beads, except for the polystyrene beads trial,
where we used polystyrene beads for both the flow and the aggregate.
The lines of slope -1 show a trend. Typical error bars error bars
estimated from error measurements are smaller than the markers. However,
experiments with similar viscosity display some variability. \label{fig:eta}}
\end{figure}

Silicone Oils allowed us to explore the influence of viscosity from
10 to 10000 mPa.s (V10-V10000), and have a conveniently constant surface
tension of 21 mN/m. On figure \ref{fig:eta}, we plotted the erosion
rate versus the viscosity. The decrease of the erosion rate spreads
over three decades, and a similar trend is observed for other trials
with other glass beads and polystyrene beads. The lines of slope of
-1 in log-log scales on this plot show a trend in agreement with the
results, even if we observe a slightly lower slope for the bigger
glass beads.

\begin{figure}
\includegraphics[scale=0.5]{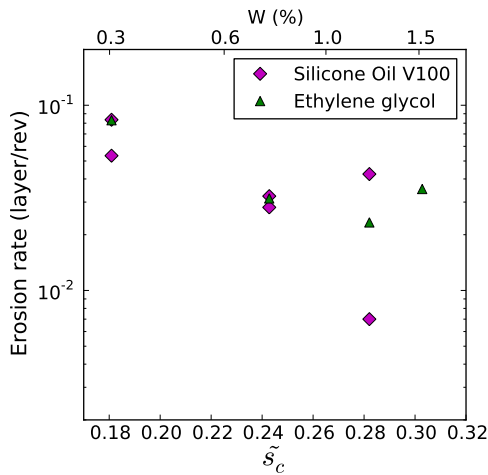}
\caption{Erosion rate plotted versus the dimensionless rupture length of the
capillary bridge. We observe similar evolution with two liquids of
different viscosity and surface tension. \label{fig:liquid content}}
\end{figure}

In the previously defined linear erosion regime, the liquid content
still has an influence on the erosion dynamics. Figure \ref{fig:liquid content}
shows the erosion rate measured for different liquid content ($W$=0.3-1.5
\%), with two different liquids. In this range, we have enough liquid
to cover the surface of the beads by a layer of liquid. Meanwhile,
we are still below a threshold of coalescence of capillary bridges,
and they are formed only between pairs of beads. Moreover, for a homogeneous
distribution the liquid network is well described in \cite{Herminghaus},
from which come the results and relations of this paragraph we rely
on. The cohesion forces have simple expressions (Eq. (\ref{eq:cohesion forces}),
and it is noticeable that theses two cohesion forces do not depend
on the volume of the bridges in first approximation. The influence
of the liquid content may be explained as follows. Under the assumption
of a homogeneous distribution of liquid, further addition of liquid
will increase the bridge volume $V$. If we assume all the liquid
to go into the capillary bridges, $V$ is linked to the liquid content
by: $\tilde{V}=\frac{8\pi W}{3\rho N}$, $V$ being rescaled by $r^{3}$,
where $N$ is the average number of bridges per bead. Therefore the
rupture length $s_{c}$ also increases as follow: $\tilde{s_{c}}=(1+\theta/2)(\tilde{V}^{1/3}+0.1\tilde{V}^{2/3})$
\cite{Willet}. $\tilde{s_{c}}$ is simply $s_{c}/r$. The rupture
length will then contribute to the effective resistance to erosion.
But more importantly, the addition of liquid also increases the connectivity
of the capillary bridges network. $N$ is experimentally given by
$N=6(1+\tilde{s_{c}})$. $\tilde{s_{c}}$, scaling roughly with $W^{\frac{1}{3}}$
(through $\tilde{V}$), appears as the relevant parameter to analyze
the liquid content influence. The equation giving $\tilde{V}$ and
$\tilde{s_{c}}$ are solved iteratively to obtain $\tilde{s_{c}}$
as a function of $W$. 

\subsection{Beads properties}

The different sets of beads used also give information about the influence
of flowing beads on the erosion dynamics. Here, only the beads of
the flow are changed, and the aggregate is always made with 200-400
$\mu$m glass beads. This way, the cohesion of the aggregate remains
unchanged, and we can independently observe the influence of the action
of the flow on the process. We changed two parameters with great consequences
on the erosion rate: bead size and bead density.

\begin{figure}
\includegraphics[scale=0.55]{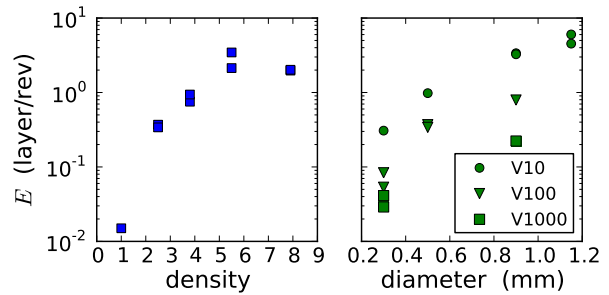}

\caption{Evolution of the erosion rate with main characteristics of flow beads.
On the left are the variations with the density of mass, using 100
mPa.s silicone oil and 500 $\mu$m beads of different material
listed in table (\ref{tab:Granular-materials-employed}). The variations
with the beads diameter are plotted on the right, using glass beads
and 10, 100 and 1000 mPa.s silicone oils. \label{fig:E diameter density}}
\end{figure}

Figure \ref{fig:E diameter density} shows a strong increase of the
erosion rate of several decades, with only moderate changes of the
mass density and diameter. The scaling appears as clearly faster than
a simple proportionality.

\section{Modeling}

\subsection{Assessment of the interactions involved}

In order to understand how the liquid properties impact the erosion
dynamics, we need to evaluate the order of magnitude of the different
forces. Since the erosion of the aggregate is a very slow process
compared to the dynamics of the flowing grains, we can assume first
that wet grains are extracted one by one from the aggregate and second,
that the dry-grains flow is not modified by this mechanism. For this
purpose, we introduce a typical velocity, $v\sim\sqrt{gd}$, which
scales as the average velocity of the first layer of beads flowing
on the aggregate. We explore a wide range of viscosity, hence the
Stokes number $St=mv/\eta d^{2}$ as defined in \cite{Ennis}, with
$m$ the mass of the beads, varies from $10^{-3}$ to $10$, which
means that inertia of the beads can be important for the low viscosities.
The capillary number was also varied in a wide range, $Ca=\eta v/\gamma\sim10^{-4}-1$.
The velocity used is relevant only if the grains are extracted at
the same speed as the flowing grains, but this is a questionable point.
Using few Fast-camera acquisitions, we measured a typical velocity
around $3$ mm/s, which is only a tenth of $\sqrt{gd}$. These measures
indicate potentially a lower capillary number, and therefore that
viscous forces are negligible versus capillary effects. This question
will be precised later.

Finally, we can evaluate the typical force $\bar{F}$ undergone by
the grains submitted to the flow. We can either choose the stress
from gravity $\tau_{g}=\rho gh\mu$, where $h$ and $\rho$ are the
height and the average density of the flowing layer and $\mu$ is
the friction coefficient at its bottom, or the one from Bagnold's
collisional-stress scaling: $\tau_{B}\propto\rho d^{2}\dot{\gamma}^{2}$,
with $\dot{\gamma}$ the shear rate in the flow. The shear rate in
the rotating drum geometry is given by $\dot{\gamma}=\frac{1}{2}\sqrt{g/d}$
\cite{GdRMiDi}, so we obtain that the stress driving the erosion
should scale either like: 
\begin{equation}
\tau_{B}\propto\rho gd\label{eq:shear stress}
\end{equation}
or, 
\begin{equation}
\tau_{g}\propto\rho gh\mu\label{eq:shear_stressgravity}
\end{equation}
The former scaling (\ref{eq:shear stress}) has been confirmed by
numerical simulations at the front of a bidisperse granular flow:
Yohannes \textit{et al.} in \cite{Yohannes} found an average boundary
stress scaling linearly with the bead size , as in Eq. (\ref{eq:shear stress}).
Then, in both cases, the density of the beads impacts directly the
boundary stress. Erosion models derived from \cite{Archard} suggest
that $E\propto\tau$. However, we can notice that a simple proportionality
between the erosion rate and $\rho$ or $d$ could not explain the
important effect on the erosion rate we observed on figure \ref{fig:E diameter density}.
This dependency would be too weak compared to the one we measured.
Then either we do not have a simple scaling of the erosion rate, or
the average shear stress is not a relevant parameter in the process.
The average force $\bar{F}$ can be evaluated with the average shear
stress (Eq. \ref{eq:shear stress} or \ref{eq:shear_stressgravity})
on one bead, $\bar{F}=\tau\pi r^{2}$, leading to a range of force
between $\bar{F}\simeq5\:10^{-7}$N for the Bagnold stress and $2.5\:10^{-6}$N
for the gravity force. We can use the Shield number $\Theta$ comparing
the tangential and confining forces, usually defined for the river
bed erosion \cite{Charru}. Here $\Theta=\bar{F}/F_{cap}$, as the
dominant confining force is $F_{cap}$. We find that $\Theta$ spans
from 0.005 up to 0.12 depending on the chosen stress and on the surface
tension. The shear force values are then at least ten times less than
capillary forces. This comparison means that erosion is not possible
to occur as a simple stretching of the bonds by the average stress.

\subsection{Stochastic approach}

Even if the average force undergone by the aggregate's grains is weak
compared to the capillary forces, physical quantities are known to
have large fluctuations around their average values in a granular
medium. Such fluctuations in the granular flow can overcome the cohesion
of the aggregate.

\subsubsection{Shear stress distribution}

Only larger values present in the force distribution $P(f)$ are able
to overcome the threshold allowing the stretching of the bonds, until
we reach the rupture. Then, in order to build an erosion rate, we
need the time for a bridge to reach the rupture, for each level of
stress. $F_{cap}$ is the static force, which defines the lower level
of erosion resistance, as the viscous force arises only with stretching
speed. We consider that only forces $f$ greater than $F_{cap}$ will
contribute to the erosion, at a rate $1/t_{rupt}$, and with a probability
of occurrence $P(f)df$. Under such hypothesis, the erosion rate derived
from this stochastic model should follow :

\begin{equation}
E_{s}=\frac{\alpha T}{2\pi}\intop_{F_{cap}}^{\infty}\frac{1}{t_{rupt}(f)}P(f)df.\label{eq: Stochastic Model}
\end{equation}
We call $E_{s}$ the theoretical erosion rate derived from this stochastic
model. $\frac{\alpha}{2\pi}$ simply represents the portion of the
aggregate undergoing the stress of the flow, and we have a dimensionless
erosion rate multiplying by $T$. Writing this model, we assume that
the rupture time is lower than the correlation time of the forces.
Gardel $\textit{et al.}$ report a correlation time of 10 ms in a
hopper for 3 mm beads \cite{Gardel}. We will comment this assumption
in the next paragraph. The Fluctuations are typically exponential
in a granular material. We chose to use the distribution derived from
the q-model, verified for static pile \cite{Coppersmith} and for
flow under shear as well \cite{Miller}:

\begin{equation}
P(f)=\frac{f^{2}}{2F_{0}^{3}}e^{\frac{-f}{F_{0}}},\label{eq:Stress distribution}
\end{equation}
$F_{0}$ being linked to the average force of the distribution by
$\bar{F}=3F_{0}$. Regardless of the precise distribution, exponential
decrease for high forces is a generic feature for granular medium
in a wide variety of conditions \cite{Mueth,Longhi}. This approach
brings us back to the previous issue of the velocity of extracted
grains : different levels of forces lead to different rupture times,
thus to various extraction velocities. From this we need to evaluate
the rupture time relations with the liquid properties as well as the
stress level.

\subsubsection{Capillary bridge dynamics}

Evaluating the rupture time requires studying the bridge dynamics.
Initially, the main force acting on the bridge is the capillary force.
Then, as the bond is stretched, viscous forces will arise. We use
a simple equation to model the bridge dynamics:

\begin{equation}
m\frac{d^{2}s}{dt^{2}}=f-2\pi\gamma r\cos\theta-\frac{3}{2}\pi r^{2}\eta\frac{1}{s}\frac{ds}{dt}\label{eq:bridge dynamics}
\end{equation}
We consider here a single capillary bridge, submitted to a constant
traction force $f$. Here we use first order expressions of the forces,
the capillary forces, for instance, actually depend on the separation
distance \cite{Herminghaus}. We consider two different limit cases
of this non-linear differential equation (\ref{eq:bridge dynamics}).
First, for low viscosity, inertia will dominate compared to the viscous
force. Neglecting this term, we can then integrate from the contact
distance $\delta$, due to roughness, to the rupture length:

\begin{equation}
t_{rupt-inert}=\sqrt{\frac{2mr\tilde{s_{c}}}{f-2\pi\gamma r\cos\theta}}\label{eq:Inertial rupture time}
\end{equation}
If viscosity is high enough, then inertia is negligible, and again
we can easily integrate without the left-hand side to obtain the rupture
time:

\begin{equation}
t_{rupt-visc}=\frac{\frac{3}{2}\pi\eta r^{2}\ln\frac{s_{c}}{\delta}}{f-2\pi\gamma r\cos\theta}\label{eq:Viscous rupture time}
\end{equation}
We evaluate these time scales for a traction force $f$ being twice
the capillary force: we find values around 0.1 ms for the inertial
time, and from 0.3 to 300 ms for the viscous time. Comparing to the
correlation time of forces from literature, the most viscous case
and the lower levels of forces will not verify the assumption we made.
Still, very large forces can achieve the rupture of the bridge in
a short enough time. To push further the analytical development of
the model, now we make the assumption that the sum of the two characteristic
times provides a good approximation of the actual rupture time.

\subsubsection{Erosion Rate}

Using the force distribution (\ref{eq:Stress distribution}) and the
total rupture time (\ref{eq:Inertial rupture time}) and (\ref{eq:Viscous rupture time}),
the erosion rate can then be developed in the following form:

\begin{equation}
E_{s}=\frac{\alpha T}{4\pi F_{0}^{3}}\intop_{F_{cap}}^{\infty}\frac{f^{2}e^{\frac{-f}{F_{0}}}}{\frac{a}{\sqrt{f-F_{cap}}}+\frac{b}{f-F_{cap}}}df
\end{equation}
with $a=\sqrt{2mr\tilde{s_{c}}}$ and $b=N_{c}\frac{3}{2}\pi\eta r^{2}\ln\frac{s_{c}}{\delta}$,
constant coefficients depending on the beads and liquid properties.
We have to consider the multiplicity of capillary bridges through
the number $N_{c}$, which is now present both in the viscous force
(in $b)$ and in the capillary force $F_{cap}$, even if we keep the
same notation. The effective number of capillary bridges per bead
in the aggregate can be reduced next to the walls. Then we overestimate
this number, but only for a small fraction of the grains, around 10\%
as we have 20 layers of grains in the width of the drum. For the beads
submitted to erosion, at the surface of the aggregate, we consider
this number to be reduced to the half: $N_{c}=N/2=3(1+\tilde{s_{c}})$.
The substitution $u=\frac{f-F_{cap}}{F_{0}}$ allows to underline
the main physical trends in the expression:

\begin{equation}
E_{s}=\frac{\alpha T}{6\pi^{2}}\,\frac{F_{cap}^{2}}{F_{0}\eta N_{c}r^{2}\ln\frac{s_{c}}{\delta}}\, e^{-\frac{F_{cap}}{F_{0}}}\: I(\frac{F_{0}}{F_{cap}},\frac{a}{b}\sqrt{F_{0}}),\label{eq:Stochastic erosion rate}
\end{equation}

with 
\begin{equation}
I(\frac{F_{0}}{F_{cap}},\frac{a}{b}\sqrt{F_{0}})=\intop_{0}^{\infty}\frac{u(1+\frac{F_{0}}{F_{cap}}u)^{2}}{1+\frac{a}{b}\sqrt{F_{0}u}}e^{-u}du.\label{eq:Integral_I}
\end{equation}

The first result of this model is the main dependencies of the erosion
rate: it decreases exponentially with the capillary force, and scales
as $\eta^{-1}$ as we observed on figure \ref{fig:eta}. The dimensionless
integral $I$ varies with dimensionless numbers as well, $\frac{F_{0}}{F_{cap}}$,
which is $\frac{3}{N_{c}}\Theta$, and $\frac{a}{b}\sqrt{F_{0}}$.
The second one can actually be written as a combination of the first
one, and other usual numbers: $\frac{a}{b}\sqrt{F_{0}}\propto\Theta\sqrt{St/Ca}$.
These numbers are rather small for the set of parameters used in most
experiments. Then $I$ is close to its limit 1, and will give weaker
influence on the erosion rate with the physical parameters. It is
worth noting that we do not expect this stochastic erosion rate to
exactly match the experimental data. Indeed, the roughness of the
surface or the local variations of the number of bridges, that would
delay the erosion, are not taken into account. 

In a different limit of $\Theta\gg1$ and $\Theta\sqrt{St/Ca}\ll1$,
the viscous rupture time would dominate, and we obtain a different
expression for the erosion rate: 
\begin{equation}
E_{s}=\frac{\alpha T}{2\pi}\intop_{F_{cap}}^{\infty}\frac{f}{b}P(f)df\simeq\frac{\alpha T}{2\pi b}\bar{F}.\label{eq:E_lowFcap}
\end{equation}
Then, the erosion rate is proportional to the average force, according
to the wear models described in \cite{Archard}.

In the following, we confront our model with the experimental data
in more detail.

\section{Analysis and discussion}

\subsection{Influence of liquid properties}

\begin{figure}
\includegraphics[scale=0.5]{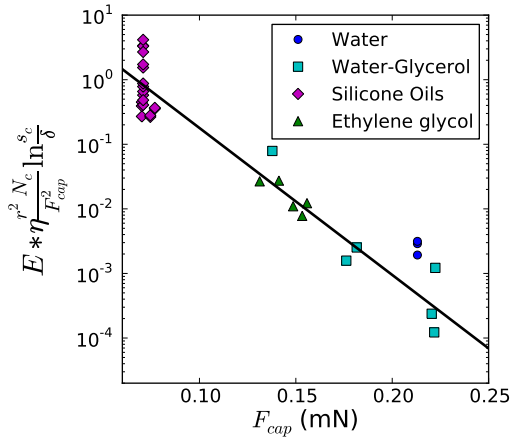}\caption{Exponential decrease of the erosion rate with the capillary force.
Experiments realized with 200-400 $\mu$m glass beads. The line
is an exponential fit of the data. \label{fig:E_Fcap} }
\end{figure}

Plotting the erosion rate times $\frac{\eta N_{c}r^{2}}{F_{cap}^{2}}\ln\frac{s_{c}}{\delta}$
(figure \ref{fig:E_Fcap}) allows to represent liquids of different
physical properties on the same graph, and should exhibit an exponential
decrease with $F_{cap}$, neglecting the variations of $I$. The points
are well gathered on the same line in log-lin scales, except for the
experiments with water. This shift is probably due to evaporation
during the experiment, reducing the effective liquid content in the
aggregate. An evaporation test showed that half the initial water
content in the aggregate disappeared after 40 minutes, which was the
time of measurement for this experiment. Similar tests on the other
liquids showed no effect of evaporation. The scaling with $\eta^{-1}$
of the erosion rate is also confirmed by this plot. The fit provides
an evaluation of the parameter $F_{0}$, related to the average force
$\bar{F}$. We found $\bar{F}=5.7\:10^{-5}$N for the standard set-up
of our experiments, that is glass beads of 200-400 $\mu$m diameter.
Nevertheless, it is straightforward to integrate numerically equation
(\ref{eq: Stochastic Model}), and fit $\bar{F}$ by successive iterations:
we find $\bar{F}=7.3\:10^{-5}$N. As this value is close to the previous
one, the first order of the variations are well captured by the first
parts of equation (\ref{eq:Stochastic erosion rate}), meaning that
the integral $I$ (Eq. \ref{eq:Integral_I}) has a moderate variation.

However, this value is quite different from the force estimation based
on dimension analysis (Eq. \ref{eq:shear stress},\ref{eq:shear_stressgravity}),
but still less than the capillary force, whose comparison justify
our stochastic approach. An explanation would be that the surface
of the aggregate is not flat and that the highest wet grains experience
larger forces, leading to a faster erosion of those grains. Another
tentative explanation is to consider capillary bridges to break one
at the time, decreasing by a factor $3(1+\tilde{s_{c}})$ the cohesive
force and the fitted $\bar{F}$. In both cases, the erosion rate is
expected to increase and to lead to an overestimation of the mean
force $\bar{F}$. Moreover, the fitted value of $\bar{F}$ is sensitive
to the precise distribution function $P(f)$, that still remains subject
to research. Nevertheless, the exponential decrease arising from the
distribution tail is a strong result unrelated to the value of $\bar{F}$.
In the next section, we test our model with the scaling of the exerted
force by the flow.

\subsection{Influence of flow properties}

The exponential decrease with the capillary force confirms an important
point of this model, which is the role of stochastic fluctuations
in the erosion process. Now as we dispose of a data set where only
the average stress is varied, we can confront the results with the
model predictions. As the aggregate is prepared likewise for each
set, using glass beads, there is no change in capillary force or in
the viscous term in equation (\ref{eq:Stochastic erosion rate}).
However, the average force, and therefore $F_{0}$, will change with
$\rho$, and $d$ according to equation (\ref{eq:shear stress}).

\begin{figure}
\includegraphics[scale=0.55]{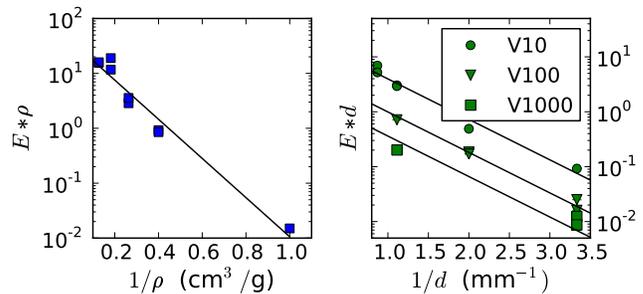} \caption{Exponential decrease of the erosion rate with the scaling parameter
of the average force, according to the model developed above. On the
left are the variations with the density of mass and a fit, and on
the right with the bead diameter. On the right plot the V10 set is
fitted, and the line reported seems also in good agreement with the
two other sets. \label{fig:exp decrease F}}
\end{figure}

Figure~\ref{fig:exp decrease F} shows the exponential decrease of
$\mbox{\ensuremath{\rho}}*E$ versus $1/\rho$ and similar variations
on the beads size. This is the expected dependency with the average
force according to equation (\ref{eq:Stochastic erosion rate}). As
in the previous part, we integrate $I$ and fit by iterations until
we obtain an evaluation of the average force, $\bar{F}$. We convert
the result to the equivalent value for the standard flow set-up (200-400
$\mu$m glass beads), according to equation (\ref{eq:shear stress}).
The data set on the mass density gives a value of $5.4\;10^{-5}$N,
and the set on the diameter with silicone oil V10 gives $5.1\:10^{-5}$N,
which are pretty concordant, and still quite close to the previous
one, using the capillary force variations. This exponential decrease
with the dimensionless number $F_{cap}/F_{0}$ is confirmed on the
two independent parameters, by separate experiments, and supports
the relevance of considering stress fluctuations in the erosion process.
This result also confirms the use of the inertial stress for the scaling
of the force.

\begin{figure}
\includegraphics[scale=0.5]{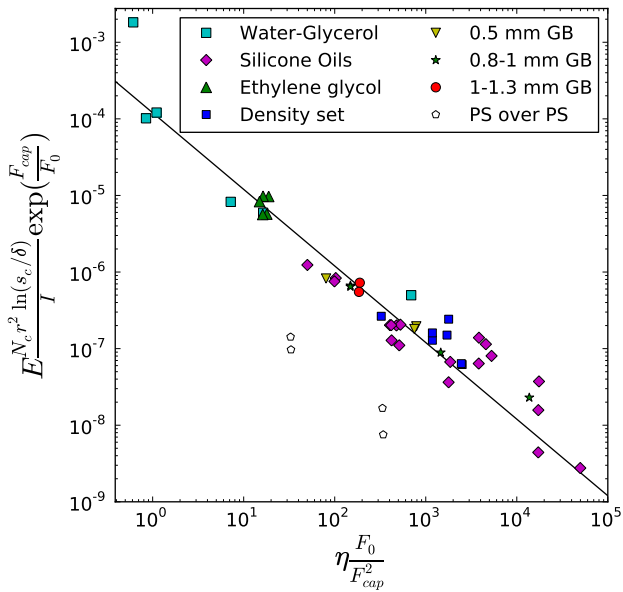} \caption{Master curve with all experimental results, enlightening the scaling
of the viscosity, extended to the different set-ups. The results with
the same x-position in a given set of data shows the variability in
the erosion rate. \label{fig:Master-curve}}
\end{figure}

The different sets of results of this erosion experiment show a good
agreement with the stochastic model proposed here. The influence on
the erosion rate of physical properties of the liquids involved (viscosity,
surface tension, liquid content), and the beads (density, size) has
been verified separately, and leads to a concordant evaluation of
the fitting parameter $\bar{F}$. Using the right scaling, we plot
all the results on the figure \ref{fig:Master-curve}, showing that
the erosion rates gather on a single master curve. The white symbols
on this figure correspond to the polystyrene beads flowing around
a polystyrene beads aggregate. They are far below the master curve
while the erosion of the polystyrene beads flowing on a glass beads
aggregate is well captured by our model (one of the darker blue square).
This is even more surprising as glass beads seem to be eroded faster
than polystyrene ones. We do not know the origin of this behavior,
and we suppose that arising static electric charges may prevent effective
contacts between beads, reducing the erosion rate. Finally, the slope
of the fitted master curve value is $\bar{F}=7.1\:10^{-5}$ N.

\section{Conclusion}

We measured the evolution of erosion rate of a wet aggregate with
respect to the liquids and grains properties. We showed an unexpected
strong influence of the surface tension. This effect is captured by
the stochastic model we proposed which shows a good agreement with
the different sets of experimental results. We cannot separate the
domain of influence of the viscosity and surface tension due to their
different roles in the erosion mechanism: we have shown that in conditions
of low Stokes number, the viscosity drives mostly the rupture time
of the capillary bridge. Meanwhile, surface tension acts as a simple
threshold and a shift for the efficient contribution in the stress
distribution of the surrounding flow. In the case of a low level of
stress, the fluctuations of the flow appear as crucial in the description
of the erosion phenomenon. This role of fluctuations have already
been pointed out for other interface behavior, expressed as boundary
conditions in \cite{Artoni2012}. Fluctuations acts also as the triggering
effect of quasistatic flows in the work of Pouliquen \textit{et al.}
\cite{Pouliquen}, similarly to our experiments, and have recently
shown their relevance in impact dynamic in granular media \cite{Clark}.

We thanks F. Chevoir for enlightening advises. We acknowledge also
P. Raux and C. Clanet for fruitful discussions on the different erosion
regimes and E. Gouillart for useful comments on the manuscript.

\end{document}